# Many Faces of Entropy or Bayesian Statistical Mechanics[*)]

E. B. Starikov[1,2]

[1)]*Institute for Materials Science and Max Bergmann Center of Biomaterials, Dresden University of Technology, D-01062 Dresden, Germany,*
*E-Mail: starikow@chemie.fu-berlin.de*

[2)]*Department of Physical Chemistry, Chalmers University of Technology, SE-412 96 Gothenburg, Sweden*

## Abstract

Some 80-90 years ago, George A. Linhart, unlike A. Einstein, P. Debye, M. Planck and W. Nernst, has managed to derive a very simple, but ultimately general mathematical formula for heat capacity *vs*. temperature from the fundamental thermodynamical principles, using what we would nowadays dub a "Bayesian approach to probability". Moreover, he has successfully applied his result to fit the experimental data for diverse substances in their solid state in a rather broad temperature range. Nevertheless, Linhart's work was undeservedly forgotten, although it does represent a valid and fresh standpoint on thermodynamics and statistical physics, which may have a significant implication for academic and applied science.

**Introduction**

Entropy is a notion which was born and successfully employed in equilibrium thermodynamics, but nowadays it turns out to be widespread and very important concept of general philosophical scale. The most recent example of this has been brought by Grisha Perelman, who used a novel "entropy-like" functional in his award-winning proof of the famous Poincaré conjecture as a part of the full geometrization conjecture [1].

But *what is* entropy ? In his famous discussion with Claude Shannon on semantic problems of the Shannon's information theory, John von Neumann is frequently quoted as saying: "… *no one knows what entropy is* …" (see, for example, [2] and the references therein). This statement still remains to be correct. Indeed, there is a wealth of entropy interpretations which are contradicting each other. Moreover, one and the same notion of entropy can be expressed using quite different mathematical functions. To this end, Peter Landsberg has once exclaimed: "*Entropies Galore !*" [3]. In accordance with this, new functional definitions of entropy are still being suggested in the literature (see, for example, [4] and the references therein).

The above is in full agreement with the earlier claim by Ed Jaynes who quoted Eugene Wigner as saying: "*Entropy is an anthropomorphic concept*" [5]. In this connection, Jaynes states: "*After the above insistence that any demonstration of the second law must involve the entropy as measured experimentally, it may come as a shock to realize that, nevertheless, thermodynamics knows of no such notion as "entropy of a physical system". Thermodynamics does have the concept of the entropy of a thermodynamic system; but a given physical system corresponds to many different thermodynamic systems*".

With this in mind, Jaynes asks: If, using one experimental set-up (or even some clever combination of different set-ups), we are exploring only a part of the whole degrees-of-freedom set of the physical system, *is there then some "true" entropy* which is a function of all these degrees of freedom ? His answer is negative: *"There is no end to this search for the ultimate "true" entropy, until we have reached the point where we control the location of each atom independently. But just at that point the notion of entropy collapses, and we are no longer talking thermodynamics !"* [5].

Still, in my opinion, it is throughout useful and instructive to look for the "ultimately true" entropy, because it was, is and will be practically impossible to *"reach the point where we control the location of each atom independently"*, as nothing in the world is a perfectly (or nearly) ideal gas. The usual way of thinking and acting here is to suggest plausible theoretical models. But these are as a rule possessed of a rather limited applicability, with tending to fail when the systems under study start to be more and more complicated. Systems' increased complexity poses generalized problems of interplay between energy and information, between micro- and macro-states, which as a rule avoid straightforward phenomenological modelling and bear philosophical significance (see, for example, [6,7] and the references therein). Hence, there is an urgent need for model-free approaches, which would at the same time be capable of taking into account all the difficulties connected with the entropy concept.

Interestingly, there is practically no true argument in the literature as to the "anthropomorphism" of entropy, to the best of my knowledge (except for Adolf Grünbaum's very specific criticism as to coarse-graining schemes forming a basis of entropy definition [8]). But, along with this, nobody, except for Ed Jaynes himself, seems to take care of how to deal *in practice* with this matter of fact. In recognition of such an intrinsically *subjective* nature of the entropy concept, Jaynes started to advocate using Bayesian statistics, when looking for the foundations of statistical thermodynamics, and the true crown of his pioneering studies was the famous generalized MAXENT principle. Accordingly, among the modern developments, several works deserve more attention.

Indeed, although the conventional statistical mechanics is „tried and true", it is well known to produce a number of serious conceptual posers (see, for example, [9] for the recent comprehensive review). First of all, the „*W*" in the famous Boltzmann-Planck formula for entropy $S = k \ln W$ is just *postulated* to be the number of microstates („complexions", as Einstein denoted them) within the given macrostate. All the reasoning which starts from this postulate is nothing but pure maths, and to check the correctness of this theoretical construction as a whole, we need to carry out all the maths first (for more detailed discussion, see [10] and the references therein).

To this end, the recent work [11] has published a handy idea to mathematically express the „*W*" through *experimentally measurable* parameters like temperature „*T*" and heat capacity „*C*". This programme is very simple at the first glance but extremely promising, because it is potentially capable of bringing the conventional statistical mechanics much closer to the experimental reality, without arbitrary modelling and conceptual difficulties. Still, the work [11] itself has not gone too far, by making very strong approximations and considering several highly idealized systems and/or limiting cases only.

Is any generalization of such an approach possible ? The answer is affirmative and has already been published about 80-90 years ago in the well-known physical-chemical journals. But these great works turn out to be clean forgotten nowadays. Refreshing them is the purpose of the present essay.

**Everything new is actually well-forgotten old**

In a series of his works [12-14], G. A. Linhart has managed to derive a general closed mathematical formula for $C_V$ vs $T$ plots from the fundamental thermodynamics principles and successfully applied this result to fit the experimental data on the $C_p$ vs $T$ plots, as well as standard entropy, for different substances in their solid state in a rather broad temperature range. From here on, we adopt $C \equiv C_V$.

The paper [12] starts with a really exciting promise to "*express the specific heat as a simple function of the absolute temperature*" after the unsuccessful attempts to do so by Einstein, Nernst, Debye and Planck. Then the author writes the following passage:

"*The rate of increase of the specific heat with the entropy of a given element or compound depends upon the probability of the randomness of the individual particles. At the absolute zero, or at the point of zero kinetic energy we are quite certain that each particle will remain in a fixed position. The probability, therefore, will be unity. At relatively high temperatures the probability of that state prevailing is very nearly zero. Now, the mathematical expression of the above statements may be assumed to be proportional to the term, $\dfrac{C_\infty - C}{C_\infty}$, which at the absolute zero is unity and at relatively high temperatures approaches zero, or $\dfrac{dC}{dS} = K\left\{\dfrac{C_\infty - C}{C_\infty}\right\}.$*"

First, the author implies here that the heat capacity $C$ can be well defined only for one and the same phase of a substance, that is, away from any phase transition which would render $C$ singular.

Second, the author implicitly suggests to *express the probabilities of macroscopic states of the matter via heat capacity*. Has he managed to track down this crucial relationship ? The answer is obviously positive. Specifically, it is mathematically proven now that, starting from the Boltzmann statistics, heat capacity could be viewed as "the sum of jumps from one energy level to another" [15]. Accordingly, the above-mentioned limiting values of heat capacity, $C_\infty$, could be viewed as the *maximum possible* sum of such "jumps" for the substance involved. Hence, the ratio $P = C/C_\infty$ could be treated as a probability that all the necessary "jumps from one energy level to another" have occurred (in other words, the corresponding macrostate has been achieved), with probability of the corresponding *disjoint event* – all the necessary "jumps from one energy level to another" have **not** occurred (that is, the corresponding macrostate has **not** been achieved) – being obviously equal to $1 - \dfrac{C}{C_\infty} \equiv \dfrac{C_\infty - C}{C_\infty} = 1 - P$, that is, just as stated by G. A. Linhart. It is important to note here that there are basically two reasons as to why the jumps among the levels may not occur: firstly, the energy difference between the levels is too high and the available energy is not enough to support such a jump, and secondly, there is in effect no need to jump, because the levels are (quasi-)degenerate. By somehow altering the environment of the substance under study (like heating it up, for example), it is possible to change *both* the population of the energy levels

*and* the energy differences among them (a very persuasive theoretical demonstration of this fact is presentend in the work [19]).

Still, the above-cited paragraph from the work [12] contains some confusion caused by the definition of the "*probability of the randomness of individual particles*": Its first sentence clearly means that such a probability is equal to nothing else than the $C/C_\infty$ quantity, but in all the further sentences the corresponding *disjoint event* is discussed and then used mathematically. When speaking of and working with thermodynamical parameters like internal energy, entropy and temperature, the advantage of omitting the vague terms like "disorder", "randomness" etc in favour of the microscopic picture of the energy levels and their populations has been thoroughly and persuasively demonstrated [16-26]. In particular, such a stanpoint considers entropy as a number of occupied energy levels in a substance, in the sense of the "storage system for energy": The larger the entropy, the more evenly is the stored energy of the substance distributed over the larger number of the occupied energy levels. Taking together the conclusions of the works [15-26], one can immediately recognize the fundamental relationship between entropy and heat capacity – the change in entropy shows *what* is changed in the physical-chemical system (the over-all "structure" of energy levels and their occupation), whereas heat capacity explains the modalities of *how* these changes can be accomplished ("jumps", that is, transitions among the energy levels).

To this end, the theoretical message of the works [12-14] can be understood by the modern reader as follows. We start out with the microscopic notion of energy levels in a substance, whose number and the separations among which are determined by the nature of the substance in question and by its surrounding as well. The particles of the substance somehow occupy these levels, and the modalities of the occupation can be described using probability theory. With this in mind, we

a) use the general representation of the experimentally measurable heat capacity as the sum of all the possible "jumps" among the energy levels, which are necessary to achieve any particular macrostate;

a) bear in mind that heat capacity of every substance should always be possessed of some characteristic limiting value.

From such a viewpoint, it is instructive to look for a mathematical relationship between the heat capacity and entropy. The works [12-14] implicitly use the above reasoning, as well as the third law of thermodynamics, to *mathematically arrive* at the following expression for entropy, which is related in a straightforward way both to the conventional Boltzmann formula, as well as to the Schrödinger-Brillouin notion of "negentropy":

$$S = \frac{C_\infty}{K}\log\left(\frac{C_\infty}{C_\infty - C}\right) = \frac{C_\infty}{K}\log\left(\frac{1}{1-P}\right) = -\frac{C_\infty}{K}\log(1-P) = k_B \log\left[\left(\frac{1}{1-P}\right)^{\frac{C_\infty}{k_B K}}\right] = k_B \log W, \quad (4)$$

where $k_B$ is the Boltzmann constant and $W$, as usual, is the number of microstates consistent with the given macrostate, whereas the "negentropy" could be expressed as (taking into account that $C_\infty \cong 3N_A k_B \equiv 3R$, where $R$ is the universal gas constant and $N_A$ – the Avogadro's number):

$$J = k_B \log\left[(1-P)^{\frac{C_\infty}{k_B K}}\right] \approx k_B \log\left[(1-P)^{\frac{3N_A}{K}}\right]. \tag{4a}$$

According to the works [12-14], *K* is the constant of the proportionality between the first derivative of heat capacity with respect to entropy and the probability that the actual macrostate has *not* been achieved. This proportionality does not seem to be directly inferable using the probability theory, but is the simplest possible mathematical manifestation of the following two physical facts [14]:

a) *both* entropy *and* heat capacity tend to decrease to zero when absolute temperature goes to zero. This is in fact a much stronger statement than merely the third law of thermodynamics – and, in effect, it is rooted in the microscopic, quantum nature of the matter [27], but can nonetheless be mathematically derived just from the macroscopic expressions for the 1$^{st}$ and the 2$^{nd}$ laws of thermodynamics without any further assumptions and/or empirical data/experience [28]. Still, it does not seem to be ever acceptable by everybody, because of the huge essential difficulty with any ultimative experimental check: It is practically impossible to achieve the absolute zero of temperature (for example, there is a detailed discussion on this topic in the works [27,29]);

a) when absolute temperature goes to infinity, entropy and heat capacity become more and more decorrelated, so that the derivative of heat capacity with respect to entropy tends to zero (this is just the reflection of the fact that heat capacity tends to its maximum value, whereas entropy does not have any limit).

A rather close resemblance among the Eq 4, the well-known Boltzmann formula and *at the time* the Schrödinger-Brillouin "negentropy" is striking – so that, it is very interesting to analyze the relationship between them. We guess that Eq 4 reflects the microscopic background for the correspondence among the macroscopic variables of the system, macroscopic constraints eventually imposed on it and the work done on or using the system (for a thorough "macroscopic" discussion on this topic, see [30]). Indeed, by removing all the possible external/internal constraints, we reduce the number of the independent macroscopic thermodynamic variables describing the system, so that the latter will reach its own most probable macrostate in accordance with the second law of thermodynamics. Imposing constraint(s), we introduce new independent macroscopic variable(s) which can be externally regulated and, in turn, ought to regulate itself the probabilities of reaching the system's own macrostates. By employing some clever schemes of imposing/removing the constraints, we are capable of causing the system to carry out some useful work for us.

The above-mentioned "imposing/removing constraints" is exactly what is dubbed "structuring" or "organizing" the system, when discussing the notion of "negentropy". The latter can be considered a "stored mobilizable energy in a space-time structured (organized) system" (see, for example, [31]), which is just a logically disjoint statement with regard to the conventional entropy as a measure of spatio-temporal "spreading of energy over all the accessible microstates in thermodynamic equilibrium" [16-25].

Bearing the above in mind, we are now conceptually ready to discuss the relationship of temperature to heat capacity, as introduced in the works [12-14], by assuming that temperature is a controllable constraint imposed on the substances under study. Using the differential form of Eq 4, G. A. Linhart transforms the conventional thermodynamical relationship $dS = (C/T)dT$ into the following differential equation:

$$\frac{dC}{dT} = \frac{K(C_\infty - C)C}{TC_\infty}, \tag{5}$$

with the straightforward solution:

$$\log\left(\frac{C}{C_\infty - C}\right) = K\log T + \log k \equiv K\log\left(\frac{T}{T_{ref}}\right) \Rightarrow \left(\frac{T}{T_{ref}}\right)^K = \frac{P}{1-P}, \tag{6}$$

where $k = 1/T_{ref}$ is the integration constant with the dimension of temperature, so that $T_{ref}$ could be some kind of a "reference temperature" dependent on the nature of the substance in question. The constants $K$, $C_\infty$ and $\log k$ are then determined by fitting Eq 6 to the experimental data on the $C_p$ vs $T$ plots, so that an excellent agreement of theory and experiment can be achieved [12-14].

The statistical interpretation of Eq 6 was not carried out in [12-14], but is now straightforward: The temperature ratio is functionally connected to the <u>*odds in favor of the reachability of the intrinsically most probable macrostate*</u>, with the $T_{ref}$ being just the temperature value where chances for a system to reach its intrinsically most probable macrostate are 50:50. Strikingly, but this is nothing else than employing the Dutch book argument, in which probabilities are defined *via* "betting odds". Remarkably, Linhart quite intuitively adopted this Bayesian-Laplacian standpoint in the times of its total rejection, long before the publication of Bruno de Finetti's and Richard Cox's groundbreaking works (Richard Cox was working on his PhD, whereas de Finetti was only about 16 years old – just in the birth year of Ed Jaynes and Peter Landsberg – when the first Linhart's paper was published) and *exactly* in the sense of the recent work on Bayesian foundations of quantum mechanics [32] ! A non-trivial step of huge importance Linhart has gone is to systematically express *both* the odds *and* the probability in the fully abstract Dutch-book argument through the *experimentally measurable* parameters, namely temperature and heat capacity, respectively. This is a perfect logical closure of his extremely simple, but fundamental, theory.

Moreover, Eq 6 represents an elegant way to redefine the conventional notion of thermodynamic equilibrium conventionally understood as a *binary* property (being/or not being in equilibrium) in terms of *continuous degrees of equilibrium*. This is not only just equivalent to using the *fuzzy set theory* (exactly in the sense as L. A. Zade defined it [33]), when speaking of thermodynamic equilibria. Importantly, as the recent work [34] shows, a redefinition of such a kind is capable of bridging a wide conceptual gap between the Boltzmann's and Gibbs' views on the equilibrium statistical mechanics, without getting into the swamp of the "ergodic theories".

Apart from all the above-mentioned and the elegant solution of the long and well known heat capacity posers [35] (as we shall see below), the practical significance of Eq 6 is that by bringing the temperature of the substance beyond its $T_{ref}$ value, we render the substance more susceptible to imposing/removing diverse constraints (in addition to the temperature) and thus capable of more efficient delivering of some useful work. For example, if we can ensure the environmental temperatures for a metallic material to stay significantly lower than its characteristic $T_{ref}$ value – and if there would be no phase transitions of the 1$^{st}$ or 2$^{nd}$ order – we might have a chance to render this material a much better electrical conductor than it would be under the normal conditions, by imposing some clever combination of further external constraints. Or, on the other hand, a metalsmith must put the corresponding metal objects into a furnace to increase their temperature and thus facilitate their forging (in the non-criminal sense, of course !).

To this end, we might refer to the $T_{ref}$ value as a kind of "critical temperature" for *a zero-th order phase transition*. Indeed, as we shall see, it can be shown that when arriving at the $T_{ref}$ value there is a singularity in the free energy, the phase of the substance in question *loses its stability* and should from that temperature point on be considered *metastable*, although entropy and heat capacity are changing continuously. We will discuss this in more detail below, but first we shall dwell on the observed correlation between the empirical parameters $K$ and log$k$ in Eq 6.

**Regression analysis of the empirical parameters $K$ and log$k$ in Eq 6**

When using Eq 6 to approximate the experimental data on molar heat capacities of solid substances in the whole range from the temperatures rather close to the absolute zero up to the corresponding melting/sublimation points, the following three parameters could have been estimated, namely $C_\infty$, $K$ and log$k$. The first parameter of these three is the maximum possible molar heat capacity of the substance, which in the works [12,13] was set to be equal to the Dulong-Petit value of $3R$ (5.966 cal K$^{-1}$ mol$^{-1}$), but, even when rendering it the full-fledged fitting parameter [14], it turns out to still remain in the nearest proximity to the $3R$ value. The deviations from the latter could be explained by the differences between $c_v$ and $c_p$, that is, molar heat capacities at constant volume and constant pressure, respectively, as well as by the experimental inaccuracy of the $C_\infty$ measurement near the melting/sublimation points. If one uses the data in Table I of the paper [14], there is practically no correlation between the $C_\infty$ and the $K$ or log$k$ values, as could be anticipated.

The practical sense of the parameters $K$ and log$k$ becomes clear, if one recasts Eq 6 in the following form [12-14]:

$$\frac{C}{C_\infty} = \frac{kT^K}{1+kT^K} = \frac{\left(\dfrac{T}{T_{ref}}\right)^K}{1+\left(\dfrac{T}{T_{ref}}\right)^K} = \frac{x^K}{1+x^K}. \qquad (7)$$

Indeed, Eqs 6 and 7 immediately suggest the method to estimate the key physical-chemical parameter $T_{ref}$ using the known values for $K$ and log$k$. To this end, it should be noted that the data on the 32 solid substances published in [12-14] are indicative of a rather strong correlation between $K$ and log$k$, even although taking together all the 32 data sets introduces some slight, but noticeable error due to the difference in the fitting approaches (sometimes with $C_\infty$ as a true fitting parameter [14], and also with the fixed $C_\infty \equiv 3R$ [12,13]). We have carefully analyzed this fundamental interdependence using linear and nonlinear regression analysis. Our main conclusion is that $K$ ought to be a polynomial of at least 3$^{rd}$ degree in log$k$ or, equivalently, in $\log_{10}((1/T_{ref})^K)$. After substitution of the resulting polynomial into the relationship $\log(k) = K\log_{10}(1/T_{ref})$ we get a handy formula to estimate the $T_{ref}$ using the fitted values of $K$ and $\log k$ for every substance:

$$K \log k = \left(Az^3 + Bz^2 + Cz + D\right)z, \quad z \equiv \log_{10}\left[\left(\frac{1}{T_{ref}}\right)^K\right]. \tag{8}$$

Along with this, the main *qualitative* result of Eq 8 is that every substance ought to have at least up to *four* different $T_{ref}$ values (more rigorously, there can be zero, two, or four real solutions to Eq 8). It is tempting to connect the latter with the characteristic temperatures of the heat capacity contributions from the four fundamental atomistic degrees of freedom of the matter, namely: translational, rotational, vibrational and electronic ones. The practical significance of this result is huge: Unlike in an ideal gas [15], it is not so easy to disentangle the heat capacity contributions in the solid state of complicated molecular solids. Nevertheless, systematical high-accuracy calorimetric studies (rows of different compounds similar to the works [12-14] – or dependence of $C_p$ vs $T$ plots for one and the same compound upon other environmental parameters, like pressure, humidity, electromagnetic fields etc.), when interpreted using Eqs 7 and 8, might be of help in this regard.

Interestingly, it is the appreciable interaction among the fundamental atomistic degrees of freedom in the solid state that could be the reason for the well-known notorious failure of the Debye model to correctly reproduce experimental heat capacities of solids at intermediate temperatures between the "ultimate" 0 K and the Dulong-Petit saturation regions. Indeed, on the one hand, the heat capacity $C_p$ estimated from the first principles using a quasiharmonic approximation (where anharmonicity approximately enters *via* a non-quadratic behavior of the internal energy *vs.* volume, as well as through a non-negligible dependence of the phonon frequencies on the volume) fits the experimental data remarkably well [36]. On the other hand, an essential anharmonicity of nuclei motion may, for example, result from a non-negligible *electron-vibrational* (vibronic) coupling (see, for example, [37]). Besides, *rotational* degrees of freedom might become appreciably anharmonic owing to rotational-vibrational coupling (see, for example, [38]). Finally, at elevated temperatures there should be a drastic increase in the *translational* contribution to the heat capacity of solids due to vacancy dynamics (see, for example, [39-41] – one might speak of vacancies themselves, as well as vacancy-atom and vacancy-vacancy interaction which would definitely give rise to the noticeable anharmonicity, as the vacancies concentration increases).

With all the above in mind, we are ready to discuss the universal temperature dependences of entropy, internal energy and Helmholtz free energies.

**Thermodynamical parameters derived from experimentally measured heat capacity. Application to supercooled and superheated bulk water.**

When considering different thermodynamical variables and functions, the conventional way of reasoning is to construct a model Hamiltonian of the system and then, with the help of this Hamiltonian, the pertinent partition function in accordance with the prescriptions of statistical mechanics/thermodynamics. From the resulting partition function the system's free energy, entropy, heat capacity etc. can be mathematically derived – but these are ultimatively model-dependent, so that such theories can be easily overthrown as a whole by the relevant experiments if the underlying model is incorrect/incomplete.

In 1999 Lieb and Yngvason have published yet another, rigorous mathematical approach to infer all the thermodynamic variables and functions – starting from entropy as a conceptual cornerstone [42,43] – and thus helping to render macroscopic equilibrium thermodynamics a logically consistent theory which is fit enough to be easily understood by the interested beginners [44].

Most recently, a noteworthy attempt has been undertaken to place macroscopic thermodynamics upon a solid system-theoretical basis using the full mathematical rigor of classical mechanics [45]. The latter approach definitely carries a big scientific potential to become a true physical theory, if it will be possible to clarify the mathematical interrelation between the moderate-sized dynamical system theory presented in the book [45] and the usual statistical thermodynamics.

The Lieb-Yngvason theory is instead much nearer to physics, in that it starts with introducing entropy as a natural measure of *adiabatic reachability* of an equilibrium macrostate in some macroscopic system. "Adiabatic reachability" means that some equilibrium state can be reachable merely on its own – that is, without carrying out work on the pertinent system. But still – entropy is not directly measurable: We can judge on its changes by measuring temperatures and comparing them to each other. In this regard, the Lieb-Yngvason theory, as it stands, still requires some *a priori* modelling to express macroscopic entropies (as well as other thermodynamical parameters and functions, including temperature itself) via statistics of the corresponding microstates for every particular case.

It is exactly the latter requirement that is perfectly avoided within the Linhart's approach. Specifically, to describe the adiabatic reachabilities, Linhart introduces probabilities of the two disjoint events – namely, whether the equilibrium macrostate is reached or not. These both disjoint events are, on the one hand, intimately connected with the heat capacity and, on the other hand, in full conformity with the Boltzmann reasoning. This is why, the Linhart's approach is basically model-free, because any *a priori* modelling of the microscopic modalities is not more necessary for the common practical usage of macroscopic thermodynamics: The microscopic-probabilistic measure of the adiabatic macrostate reachability could be directly extracted from the systematic experimental results on heat capacity changes vs. different external factors (with the temperature being always among those). After we have thus learned about <u>how and to which extent</u> the macroscopic states are reached in the real life, we might wish to judge on their *general reachability* (that is, on their entropies, using the Lieb-Yngvason theory [42]), and/or their *general controllability* (using the system-theoretical approach [45]).

Furthermore, Linhart's approach allows to go even one further step forward by naturally introducing (through the probabilities of disjoint events) the parameter dual to entropy and known as the Schrödinger-Brillouin "negentropy". Interestingly, the novel dynamical theory of thermodynamics [45] also defines an analogous parameter, which is dubbed "ectropy" – but the definition of the latter is rather artificial, it is simply a kind of "gift from heaven" without any physical explanation as to *why* the ectropy *must* always accompany entropy.

With all the above in mind, we shall now go the *reversed* way – as compared to the conventional statistical thermodynamics – to analyze all the well-known thermodynamic variables and functions starting from the heat capacity defined by Eq 7. Actually, as pointed out in [12], Eq 7 is better suited to describe heat capacities at constant volumes, $C_V$, than heat capacities at constant pressures, $C_p$. It was empirically and computationally proven [12-14,35,36] that the former is in general a good approximation to the latter at low and intermediate temperatures, but at higher temperatures there are always discrepancies between the both, because $C_V$ gets saturated at approximately Dulong-Petit level, whereas $C_p$ is still noticeably temperature-dependent. But for the sake of completeness we would first like to deal with the isochoric-isothermal case, before discussing the differences between $C_V$ and $C_p$ in detail and working with isobaric-isothermal systems. Indeed, it is straightforward to derive the expressions for the temperature dependences of entropy $S(T)$, internal energy $U(T)$ and Helmholtz free energy $F(T)$ as follows (for $K > 0$), using Eq 7:

$$S(T) = \int_0^T \frac{C_V(T)}{T} dT = C_\infty \int_0^x \frac{x^{K-1}}{1+x^K} dx = \frac{C_\infty}{K} \ln(1+x^K), \tag{9}$$

$$U(T) = \int_0^T C_V(T) dT = T_{ref} C_\infty \int_0^x \frac{x^K}{1+x^K} dx = C_\infty T_{ref} x \left[1 - {}_2F_1\left(1, \frac{1}{K}, 1+\frac{1}{K}; -x^K\right)\right], \tag{10}$$

$$F(T) = U(T) - TS(T) = C_\infty T_{ref} x \left[1 - {}_2F_1\left(1, \frac{1}{K}, 1+\frac{1}{K}; -x^K\right) - \frac{1}{K}\ln(1+x^K)\right], \tag{11}$$

where ${}_2F_1(a, b, c; z)$ stands for the conventional Gauss hypergeometric function.

First, we would like to investigate mathematical behavior of the thermodynamical functions under study at one particular $T = T_{ref}$ or, equivalently, $x = 1$ point. As the Gauss hypergeometric function is real only for the region $|x| \leq 1$, we need a separate investigation of the definite integrals in Eqs 9, 10, but now arriving at the same temperature $T$ from the plus-infinity. As a result, we have (for the real parameters $K > 0$ and $x = T/T_{ref}$):

$$\widetilde{S}(T) = \int_\infty^T \frac{C_V(T)}{T} dT = -C_\infty \int_x^\infty \frac{x^{K-1}}{1+x^K} dx = C_\infty \int_\infty^x \frac{x^{K-1}}{1+x^K} dx = -\infty, \tag{12}$$

$$\widetilde{U}(T) = \int_\infty^T C_V(T) dT = T_{ref} C_\infty \int_\infty^x \frac{x^K}{1+x^K} dx = C_\infty T_{ref} x \left[{}_2F_1\left(1, -\frac{1}{K}, 1-\frac{1}{K}; -x^{-K}\right)\right], \tag{13}$$

$$\widetilde{F}(T) = \widetilde{U}(T) - T\widetilde{S}(T) = C_\infty T_{ref} x \left[{}_2F_1\left(1, -\frac{1}{K}, 1-\frac{1}{K}; -x^{-K}\right) - \frac{1}{K}\ln(1+x^K)\right]. \tag{14}$$

Now, finding the $x \to 1$ limits of Eqs 7, 9-11, 13, 14, we see that, whereas $C_V(T)$ and $S(T)$ are continuous at $T = T_{ref}$, $U(T)$ and, accordingly, $F(T)$ exhibit a discontinuity, the nature of which is strongly dependent on the magnitude of $K$. This is why, we may consider *any* real value of $T_{ref}$ a threshold for a "hidden" zeroth-order phase transition. If there are two or four possible values of the $T_{ref}$ along the whole temperature range, one should respectively have two or four discontinuities in $U(T)$ and $F(T)$.

Interestingly, zero-order phase transitions are not "science fiction": they could in principle be the consequence of long-range interactions in a system [46] and have definitely been shown to underlie the solid-solid volume collapse transitions in cerium and related compounds [47]. Although it should be noted that when speaking of "zero-order phase transitions" one sometimes understands singularities not only in the free energy, but in entropy and heat capacity at the same time [48].

Further, it is also important to note the correspondence between the Debye theory and the Linhart approach. Indeed, one can readily check by direct substitution [49] that the well-known integral Debye functions describing $C_v(T)$ and $U(T)$ of a $n$-dimensional crystal obey the following functional identity (which mathematically is nothing else than just the Legendre transformation):

$$DC_n(x) = xDE_n(x) - x\frac{d}{dx}[xDE_n(x)],$$

$$DE_n(x) \equiv \frac{n}{x^{n+1}}\int_0^x \frac{t^n dt}{\exp t - 1}; \quad DC_n(x) \equiv \frac{n}{x^n}\int_0^x \frac{t^{n+1}\exp(-t)dt}{[1-\exp(-t)]^2}; \quad (15)$$

$$x \equiv \frac{T_D}{T}; \quad U(T) \equiv DE_n\left(\frac{T_D}{T}\right); \quad C_V(T) \equiv DC_n\left(\frac{T_D}{T}\right); \quad T_D \equiv \text{Debye temperature}.$$

Remarkably, it is straightforward to demonstrate that the same Legendre transformation identity holds with the Linhart expressions for $C_v(T)$ (Eq 7) and $U(T)$ (Eq 10) as well. This helps to find by the analogy the physical sense of the parameter $K$ in Eq 7: It ought to be nothing else than *geometrical dimension of the sample*. Indeed, the fundamental difference between the Debye and Linhart models consists in that the former describes idealized extended systems with *integer* spatial dimensions, whereas the latter provides a theoretical basis for the extended systems with *fractal* dimensions, because $K$ in Eqs 7–14 is a *real* number, while $n$ in Eq 15 is always *integer*.

Meanwhile, the exact and rigorous mathematical correspondence between the Debye and Linhart approaches should definitely be non-trivial, because the scalar parameter $T_D$ in the Debye model corresponds to a set of $T_{ref}$ values in the Linhart's approach. Along with this, the Debye model of a solid considers solely vibrational degrees of freedom in the form of linear harmonic oscillators, whereas the Linhart's approach ought to contain all the possible microscopic degrees of freedom of the condensed matter, as discussed above. Hence, Linhart's theory can be considered a non-trivial generalization of the Debye's theory of an ideal solid for the cases of realistic extended condensed-matter systems with all the possible kinds of disorder leading to fractalization of the conventional integer spatial dimensions.

It is now interesting to look for the exact mathematical expression of the probability measure corresponding to the Linhart's theory. As the latter has only temperature and geometrical dimension as its parameters, it is straightforward to use the geometrical analogy of thermodynamics as developed in the works [50,51], where thermodynamical formalism is employed to study intersections of a generalized plane curve with a random line. Specifically, we start with the expression for the entropy in Eq 9 and consider its Taylor expansion in the powers of $x^K$, which is convergent when $|x^K| < 1$. We immediately have

$$\log(1+x^K) = -\sum_{k=1}^{\infty} \frac{(-1)^k x^{kK}}{k}, \tag{16}$$

and just comparing this series with the Shannon-like "geometrical" entropy introduced in [50,51]:

$$S_m = -\sum_{k=1}^{\infty} m_k \log m_k, \tag{17}$$

where $m_k$ is the probability that all the $k$ random lines will intersect the given plane curve – or the desired statistical-mechanical probability measure, we finally obtain that

$$m_k = \frac{W(z_k)}{z_k}; \quad z_k = \frac{(-1)^{k+1} x^{kK}}{k}, \tag{18}$$

where $W(z)$ is the Lambert $W$-function [52]. Interestingly, the Linhart's $m_k$ is possessed of just the same functional form, as those derived for the Tsallis' entropy expressions using the Jaynes' MAXENT approach [4,53]. As we have not employed the MAXENT approach here, Eqs 16-18 prove that the Linhart's theory should be consistent with the Tsallis generalization of the Boltzmann-Gibbs statistical mechanics.

Finally, we would like just to show a way to derive the equation of state corresponding to the Linhart's theory. We shall make all the usual assumptions about the statistical-mechanical system under study (see, for example, [54]) and define its configurational integral and concentration activity coefficient $\gamma$. The latter can be regarded as a function of temperature and the density $\rho = N/V$, where $N$ stands for the number of particles in the system and $V$ is its volume. Meanwhile, $V$ and, accordingly, $\rho$ are themselves functions of temperature, so, introducing the effective temperature $x$ as in Eq 7, we may express the Helmholtz free energy per one particle of a non-ideal gas via $\gamma$, assuming that all the temperature dependence of $\gamma$ comes from the density only:

$$\frac{F(x)}{k_B T_{ref}} = -x\rho^{-1}(x)\int_0^\rho \ln\gamma(\tilde{\rho})d\tilde{\rho} = x\rho^{-1}(x)\int_x^\infty \ln\gamma(X)\frac{d\rho(X)}{dX}dX, \tag{19}$$

where $k_B$ is Boltzmann constant. Let $\rho = 0$ be corresponding to $x = \infty$ (that is, there is thermal expansion, as usual, although the rare cases of thermal contraction can also be handled the proper way). Then, it is possible to integrate Eq 19 by parts, and we recast it as follows:

$$\rho^{-1}(x)\int_x^\infty \ln\gamma(X)\frac{d\rho}{dX}dX = \ln\gamma(x) - \rho^{-1}(x)\int_x^\infty \rho(X)\frac{(d\gamma(X)/dX)}{\gamma(X)}dX. \tag{20}$$

On the one hand, it is comparable with Eq 11, so that we have

$$\gamma(X) = (1+x^K)^{-\frac{C_\infty T_{ref}}{K}}; \quad -\frac{(d\gamma(X)/dX)}{\gamma(X)} = C_\infty T_{ref}\frac{X^{K-1}}{1+X^K} \tag{21a}$$

and

$$\rho^{-1}(x)\int_x^\infty \rho(X)\frac{X^{K-1}}{1+X^K}dX = 1 - {}_2F_1\left(1, \frac{1}{K}, 1+\frac{1}{K}; -x^K\right). \tag{21b}$$

On the other hand, the equation of state can be written as follows [54]:

$$\frac{P(\rho,x)}{k_B T_{ref}} = x\rho\left[1 + \rho^{-1}\int_0^\rho d\tilde{\rho}\tilde{\rho}\frac{\partial \ln\gamma(\tilde{\rho},x)}{\partial \tilde{\rho}}\right], \tag{22}$$

where $P(\rho,x)$ is pressure. To get the exact mathematical expressions for the temperature dependences of $P$ and $\rho$, we have to solve the integral-functional equation Eq 21b with respect to $\rho(x)$ and subsiture the result into Eq 22, but we will not dwell on these maths here.

**Conclusions**

1. In his clean-forgotten works, George A. Linhart has suggested theoretical basis for the "Bayesian statistical mechanics", long before the publications by Ed Jaynes.

2. Linhart has shown, how the famous Boltzmann-Planck formula for entropy can be *mathematically derived* starting from experimental facts (and not just postulated, as done usually).

3. The entropy formula derivation by Linhart also demonstrates the intrinsic connection between the entropy and Schrödinger-Brillouin "negentropy" notions.

4. Linhart's approach allows to bridge a logical gap between the Boltzmann and Gibbs interpretations of statistical mechanics, without applying to "ergodic theories", but reconsidering the very notion of "thermodynamical equilibrium".

5. Pursuing the Linhart's line of thought, it is possible to show not only mathematical, but also *physical* interconnection between the Boltzmann-Gibbs and Tsallis statistics.

6. Linhart's ideas pave the way of mathematical derivation of the universal equation of state.